\documentclass[aps,pra,notitlepage,preprint,nofootinbib,eqsecnum,showkeys,%
showpacs,groupedaddress,a4paper,byrevtex]{revtex4-1}
\usepackage{bm}
\usepackage{amsmath}
\usepackage{amsfonts,amssymb}
\begin{document}
\newcommand{\W}{\boldsymbol{\mathfrak{W}}}
\newcommand{\rmi}{\text{i}}
\newcommand{\tm}{T^{\text{M}}_{\alpha \beta}}
\newcommand{\ta}{T^{\text{A}}_{\alpha \beta}}
\newcommand{\tsym}{T^{\text{\,sym}}_{\alpha \beta}}
\newcommand{\g}{\ensuremath{\overline\Gamma}}
\newcommand{\wt}{\widetilde{\Omega}}
\title{Symmetric energy-momentum tensor: \\ the Abraham form
and the explicitly covariant formula}
\author{V.V.~Nesterenko}
\email[E-mail:~]{nestr@theor.jinr.ru}
\author{A.V.~Nesterenko}
\affiliation{Bogoliubov Laboratory of Theoretical Physics,
Joint Institute for Nuclear Research, \\
141980 Dubna, Russia}
\begin{abstract}
We compare the known in literature, explicitly covariant 4-dimensional
formula for the symmetric energy-momentum tensor of electromagnetic field
in a medium and the energy-momentum tensor derived by Abraham in the
3-dimensional vector form. It is shown that these two objects coincide
only  on the physical configuration space~$\overline\Gamma$, formed by the
field vectors and the velocity of the medium, which satisfy the Minkowski
constitutive relations. It should be emphasized  that the 3-dimensional
vector formulae for the components of  the energy-momentum tensor were
obtained by Abraham only on~$\overline\Gamma$, and the task of their
extension to the whole unconditional configuration space~$\Gamma$ was not
posed. In order to accomplish the comparison noted above we derive the
covariant formula anew by another method, namely, by generalizing the
Abraham reasoning. The comparison conducted enables one to treat the
explicitly covariant formula as a unique consistent extension of the
Abraham formulae to the whole configuration space $\Gamma$. Thus the
question concerning the relativistic covariance of the original
3-dimensional Abraham formulae defined on $\overline\Gamma$ is solved
positively. We discuss in detail  the relativistic covariance of the
3-dimensional vector formulae for individual components of the
4-dimensional tensors in electrodynamics which is manifested in the
form-invariance of these formulae under Lorentz transformations.
\end{abstract}

\maketitle
\newpage
\tableofcontents
\newpage
\section{Introduction}
In the electrodynamics of continuous media, it is generally accepted that
the symmetric energy-momentum tensor of electromagnetic field was derived
by Abraham (see, for example, the comprehensive survey by
Pauli \cite{Pauli}. However, it is worth noting that Abraham has derived
the components of his energy-momentum tensor only assuming that the
Minkowski  constitutive relations are satisfied, i.e., only on the
physical configuration space $\g$ of changing the field vectors
$\mathbf{E,\;H,\;D,\;B}$ and the velocity of the medium $\mathbf{v}, \;
v<c$. The task to extend the Abraham formulae to the whole unconditional
configuration space~$\Gamma$ was not posed. This peculiarity of the
Abraham formulae completely concerns also the \mbox{4-dimensional}
presentation of these formulae proposed by Grammel \cite{Grammel}. This
fact explains in particular the lack of uniqueness of the Grammel formulae
(three versions) and their explicit dependence  on $\varepsilon$
and~$\mu$. Obviously, the Abraham and Grammel formulae should be used only
on $\g$, where they coincide and do not depend explicitly on $\varepsilon$
and~$\mu$, but not on the whole configuration space $\Gamma$. In the Pauli
survey \cite{Pauli} and in the subsequent papers dealing with the Abraham
and Grammel formulae these peculiarities were not noted and not taken into
account~\cite{Leonhardt,Ginzburg-Ugarov,Skob,Brevik-1,Brevik-2,Ginzburg-73}.

Let us make more precise the used terminology. Following Pauli, we define
the symmetric energy-momentum tensor by its known values in co-moving
frame \cite[Eqs.~(298) and~(300)]{Pauli}. By~Abraham's formulae, or simply
by the Abraham tensor, we call, following again Pauli \cite{Pauli}, the set
of formulae for all the components of the energy-momentum tensor, which
were obtained by Abraham on $\g$ (see \S\S38,~39 in his book
\cite{Abraham1920} and also in his articles \cite{AP1909, AP1910}). These
formulae will also be brought about  in the present paper in
Section~\ref{ta}. The formulae derived by Abraham are  valid only on $\g$,
therefore in advance it is not apparently that they define the components
of a 4-dimensional relativistic vector. This point is also investigated in
our paper.

The symmetric energy-momentum tensor, as it was specified
by Pauli and defined on the whole configuration space
$\Gamma$,  has been constructed considerably later than the
Abraham works \cite{Schmutzer,Groot,Obukhov,MR}. These
authors proposed the explicitly covariant 4-dimensional
formula for this tensor, which is applicable both in the
rest frame and for moving media. It is important that this
formula does not contain explicitly the material
characteristics $\varepsilon$ and $\mu$ of the medium. In
the books \cite{Schmutzer, Groot} the covariant formula is
given without derivation. In Ref.\ \cite{Obukhov} the
symmetric energy momentum tensor was derived by
``abrahamization'' of the Minkowski tensor. Further
development and application of this method see
in \cite{ni,nii,niii}. The technique  used in \cite{MR} will be
discussed in Section \ref{f-ib} of the present paper. The
authors of all these works call their tensor by the Abraham
tensor. However, the comparison of these two objects was
not conducted. In the preset paper this gap will be filled
up.

For this aim, we generalize the Abraham reasoning which he
followed when deriving his energy-momentum tensor on $\g$.
This enables us to construct the symmetric energy-momentum
tensor on the whole configuration space $\Gamma$ in another
way different from the calculations\footnote{The method
used in \cite{MR} is discussed in Section \ref{f-ib} of the
present paper.} in papers \cite{Obukhov,MR}. In our
approach, the identity  of the symmetric energy-momentum
tensor and the Abraham tensor on $\g$ is proved
easily.\footnote{To do the same proceeding from the formula
obtained in \cite{Obukhov,MR} is a formidable task.}
In~addition, we derive at the same time the explicit
3-dimensional vector formulae for all the components of the
symmetric energy-momentum tensor. The latter is important
due to the following.

In practical calculations, one uses, as a rule, the individual components
of the energy-momentum tensor in 3-dimensional vector notation. In
particular, such a situation takes place in the electrodynamics of moving
bodies. The point is that the Lorentz vector describing the velocity of
the medium
\begin{equation}
\label{e0} u_\alpha=\gamma\{\mathbf {q},
\text{i}\},\quad  \mathbf {q}= \mathbf{v}/c,\quad
\gamma^{-1}=\sqrt{1-\mathbf{q}^2}, \quad u_\alpha
u_\alpha=-1, \quad \alpha =1,2,3,4
\end{equation}
cannot be itself considered  as a small parameter. Only its spatial part,
$\mathbf{q}$, may be such a parameter. The 3-dimensional vector formulae
for the energy-momentum tensor  in the Minkowski form  and in the Abraham
form are known~\cite{Pauli,Abraham1920}. In the present paper, the
analogous  formulae will be derived for the symmetric energy-momentum
tensor. A special attention will be paid to the discussion of the
relativistic covariance of these formulae.

The layout of the paper is as follows. In Section \ref{tm}, the formulae
are brought which define  the energy-momentum tensor in the Minkowski form
and the Minkowski constitutive relations are presented. This material is
substantially used in the following. In Section \ref{ta}, the derivation
of the energy-momentum  tensor by Abraham is traced in detail. In Section
\ref{symt} the generalization of the Abraham reasoning enables us to
construct the symmetric energy-momentum tensor  on the whole configuration
space $\Gamma$. Here the identity of the symmetric energy-momentum tensor
and the Abraham tensor on $\g$ is proved. At the same time, the explicit
3-dimensional vector formulae   are derived for all the components  of the
symmetric energy-momentum tensor. These formulae are analogous to those
derived by Abraham. In Section \ref{f-i}, we discuss different methods of
defining  the tensors of the second  rank in electrodynamics and the
form-invariance of the pertinent 3-dimensional vector formulae. It~is
shown in detail in what way this form-invariance becomes apparent in the
case of the energy-momentum tensor in the Minkowski form, in the Abraham
form, and in the case of the symmetric energy-momentum tensor. In
Conclusion, Section \ref{cncl}, the obtained results are summarized
briefly and some comments  are made.

We do not touch the Abraham-Minkowski controversy and the problem of
determining the ``correct'' energy-momentum tensor in macroscopic
electrodynamics. There is a large body of the literature on this subject,
see, for example, the reviews \cite{Brevik-1, Brevik-2, Skob, Ginzburg-73,
Ginzburg-Ugarov, MR, Obukhov} and references therein.

We use the following notations. The Greek indices take values $1,2,3,4$,
whereas the Latin indices assume the values $x,y,z$. The unrationalized
Gaussian units are used for the electromagnetic field and the notations
generally accepted in macroscopic electrodynamics \cite{LL8} are adopted.

\section{The Minkowski energy-momentum tensor}
\label{tm}

We will frequently refer to the energy-momentum tensor in the Minkowski
form. Therefore, we bring here the formulae needed. This tensor is a
straightforward generalization of the energy-momentum tensor of
electromagnetic field in vacuum~\cite{LL2},
\begin{equation}
\label{e2-1} T_{\alpha
\beta}=T_{\beta \alpha}=\frac{1}{4\pi}\left ( F_{\alpha
\gamma}F_{\gamma \beta}-\frac{\delta_{\alpha
\beta}}{4} F_{\gamma \delta }F_{\delta \gamma} \right
){,}
\end{equation}
where  $F_{\alpha \beta}=-F_{\beta \alpha}$ is
electromagnetic field tensor in vacuum,
$F_{ij}=\varepsilon_{ijk}H_k$, $\;F_{4j}=-F_{j4}=\rmi
E_{j}${.}

Minkowski \cite{Pauli,LL8} has formulated relativistic macroscopic
electrodynamics in terms of two tensors which are defined by the formulae
\begin{eqnarray}
F_{ij}&=&\varepsilon_{ijk}B_k,\quad F_{4j}=-F_{j4}=\rmi E_{j}\,{;}  \label{e2-2}\\
H_{ij}&=&\varepsilon_{ijk}H_k,\quad
H_{4j}=-H_{j4}=\rmi D_{j}\,{.} \label{e2-3}
\end{eqnarray}
Generalizing (\ref{e2-1}), Minkowski proposed the following
energy-momentum tensor of electromagnetic field in a
material medium~\cite{Pauli}:
\begin{equation}
\label{e2-4} T^{\text{M}}_{\alpha
\beta}=\frac{1}{4\pi}\left ( F_{\alpha
\gamma}H_{\gamma \beta}-\frac{\delta_{\alpha
\beta}}{4} F_{\gamma \nu }H_{\nu \gamma} \right
){.}
\end{equation}
This definition holds both for the medium at rest and for the moving
medium. According to construction, the velocity of the medium does not
enter into Eq.~\eqref{e2-4}. The Minkowski energy-momentum tensor is not
symmetric $\tm \neq T^{\text{M}}_{\beta \alpha}$ and its trace vanishes
$T^{\text{M}}_{\alpha \alpha}=0$.

By making use of the 3-dimensional vector notation we can present the
components of the tensor $\tm$ (\ref{e2-4}), in the arbitrary inertial
reference frame, in the form~\cite{Pauli}
\begin{equation}
\label{e2-5} \tm=\left (
\begin{matrix}
\sigma^{\text{M}}_{ij}&-\rmi c\,
\mathbf{g}^{\text{M}}\cr -\frac{\displaystyle
\rmi}{\displaystyle  c}
\,\mathbf{S}^{\text{M}}&w^{\text{M}}\cr
\end{matrix}
\right ){,}
\end{equation}
where
\begin{gather}
\sigma_{ij}^{\text{M}}=\frac{1}{4 \pi}\left \{
E_iD_j+H_iB_j-\frac{\delta_{ij}}{2}\,(\mathbf{ED+HB})
\right \}{,}\label{e2-6} \\
\frac{1}{c}\,
 \mathbf {S}^{\text{M}}=\frac{1}{c}\,\mathbf{S}
=\frac{1}{4\pi}\,[\mathbf{EH}]\,{,}\quad c\mathbf {g}^{\text{M}}=
\frac{1}{4\pi}\,[\mathbf{DB}]\,{,}
\label{e2-7} \\
w^{\text{M}}=\frac{1}{8
\pi}\,(\mathbf{ED+HB})\,{.}\label{e8}
\end{gather}
Equation (\ref{e2-7}) involves the Poynting vector
\begin{equation}
\label{Poynting}
\mathbf{S}=\frac{c}{4\pi}\,[\mathbf{EH}]\,{.}
\end{equation}

In macroscopic electrodynamics \cite{Pauli, LL8}, the important role is
played by the Minkowski constitutive relations:
\begin{eqnarray}
D_\alpha&=&\varepsilon E_{\alpha}\, {,} \label{Minkowski-1}\\
B_{\alpha \beta \gamma}&=&\mu H_{\alpha\beta \gamma}
\,{.}\label{Minkowski-2}
\end{eqnarray}
We use the notation from~\cite{Kafka}
\begin{eqnarray}
E_\alpha\equiv F_{\alpha \beta}u_\beta=\gamma\{
\mathbf{E}+[\mathbf{q}\mathbf{B}],\;\rmi(\mathbf{q}\mathbf{E})
\}\,{,}&\quad& D_\alpha \equiv H_{\alpha \beta}u_{\beta}=\gamma\{
\mathbf{D}+[\mathbf{q}\mathbf{H}],\;\rmi(\mathbf{q}\mathbf{D})
\}\,{,}
\label{Kafka-1} \\
B_{\alpha \beta \gamma}= F_{\alpha \beta}u_\gamma+F_{\beta \gamma}u_\alpha +
F_{ \gamma \alpha}u_\beta \,{,}&\quad&
H_{\alpha \beta \gamma}= H_{\alpha \beta}u_\gamma+H_{\beta \gamma}u_\alpha +
H_{ \gamma \alpha}u_\beta \,{,}\label{Kafka-2}
\end{eqnarray}
where $u_\alpha$ is the four-vector of the medium velocity~\eqref{e0}. In
terms of the 3-dimensional vector notation Eqs.~\eqref{Minkowski-1},
\eqref{Minkowski-2} acquire the form~\cite{Pauli, LL8}
\begin{gather}
\mathbf{D}+[\mathbf{q} \mathbf{H}]=\varepsilon\left (
\mathbf{E}+[\mathbf{q} \mathbf{B}]
\right ),\label{Minkowski-3}\\
\mathbf{B}-[\mathbf{q}\mathbf{E}]=\mu\left (
\mathbf{H}-[\mathbf{q}\mathbf{D}]
\right ){.}\label{Minkowski-4}
\end{gather}

Only due to the constitutive relations (\ref{Minkowski-1}),
(\ref{Minkowski-2}) or (\ref{Minkowski-3}), (\ref{Minkowski-4}) the formal
scheme of the macroscopic electrodynamics acquires the physical
content \cite[\S~33]{Pauli}. The values of the vectors $\mathbf{E,\; H,\;D,\; B}$
and the velocity of the medium $\mathbf{v}, \; v<c$, which satisfy the
Minkowski constitutive relations (\ref{Minkowski-1}), (\ref{Minkowski-2})
will be referred to as {\it the physical configuration space}
$\overline\Gamma$.

In the relativistic electrodynamics of moving media, since the Minkowski
works, the \mbox{4-dimensional} generalization of the Poynting vector
(\ref{Poynting}) is used (the so-called Ruhstrahlvektor~\cite{Pauli})
\begin{equation}
\label{Ruh}
\Omega_\alpha=H_{\alpha \beta \gamma}u_{\beta} E_{\gamma}\,{.}
\end{equation}
For simplicity, we shall call $\Omega_\alpha$ the Minkowski vector. In the
co-moving reference frame ($K'$), Eq.~(\ref{Ruh}) yields~\cite{Pauli,Kafka}
\begin{equation}
\label{Ruh-1}
\frac{c}{4\pi}\,(\Omega'_1,\Omega'_2,\Omega'_3)=\mathbf{S'},\quad
\Omega'_4=0\,{.}
\end{equation}
Following Pauli \cite{Pauli}, we mark quantities in a co-moving frame by
prime. From Eqs.~(\ref{Ruh-1}) and~(\ref{e0}) it follows
\begin{equation}
\label{Ruh-2}
u_\alpha\Omega_\alpha  =0\,{.}
\end{equation}

\section{The Abraham energy-momentum tensor}
\label{ta}

Abraham imposes the following conditions on the energy-momentum tensor to
be found. This tensor should be symmetric; in the co-moving reference
frame ($\mathbf{v}=0$) and under fulfillment of the constitutive relations\footnote{The account of the constitutive relations in Abraham's
derivation of the energy-momentum tensor is not noted in the Pauli
review \cite{Pauli} and in the subsequent papers, see, for example \cite{MR,Ginzburg-Ugarov, Skob, Brevik-1, Brevik-2, Ginzburg-73}.}
\begin{equation}
\label{e3-1}
\mathbf{D}=\varepsilon \mathbf{E},\quad \mathbf{B}=\mu \mathbf{H}
\end{equation}
(isotropic medium) the components of this tensor should assume the values
\begin{gather}
\label{e3-2}
\sigma^{'\text{A}}_{ij}=\cfrac{1}{2}\,(\sigma^{'\text{M}}_{ij}
+\sigma^{'\text{M}}_{ji})\,{,} \\
\label{e3-3}
\cfrac{1}{c}\,\mathbf{S}^{'\text{A}}=c\mathbf{g}^{'\text{A}}=\cfrac{1}{c}\,\mathbf{S}'
=\cfrac{1}{4\pi}\, [\mathbf{E'H'}]\,{,}
 \\
\label{e3-4}
w^{'\text{A}}=\frac{1}{8\pi}\,(\mathbf{E'D'+H'B'})\,{.}
\end{gather}

We have to stress here that all these requirements are not sufficient to
uniquely define the energy-momentum tensor on the whole configuration
space $\Gamma$. Indeed, Abraham, having been fixed the reference frame
($\mathbf{v}=0$), assigns the components of his tensor only on the
physical configuration space $\overline \Gamma$ (i.e., under fulfillment
of the constitutive relations (\ref{e3-1})), but not on the whole
configuration space $\Gamma$, as the theory of relativity demands.

However, Abraham  did not use the theory of relativity in constructing his
energy-momentum tensor. ``He went along the way of extrapolations
proceeding from the medium at rest, relying on rather arbitrary
assumptions and very poor experimental data'' \cite[p.\ 289, Russion eddition]{Skob}). Following this way, Abraham has
derived the energy-momentum tensor first in the framework of his approach
to the electrodynamics of moving bodies \cite[\S\S~38,~39]{AP1909}  and
later on in application to the Minkowski electrodynamics \cite{AP1910}.
In~the paper \cite{AP1910} Abraham approached the most close to using the
tensor formalism in the problem under consideration. The Abraham reasoning
\cite{AP1910} was ``translated'' to the standard tensor language by
Grammel \cite{AP1910}, the pertinent detailed calculations were also
conducted by Kafka \cite{Kafka}. We shall follow the papers \cite{AP1910,Grammel, Kafka}.

By making use of the standard tensor notation the Abraham reasoning
\cite{AP1910} can be reformulated like this. In the co-moving reference
frame ($\mathbf{v}=0$) and when the constitutive relations~(\ref{e3-1})
hold, the symmetric tensor
\begin{equation}
\label{ta-1}
\frac{1}{2}(\tm+T^{\text{M}}_{\beta \alpha })
\end{equation}
gives the following value for the density of the energy flux:
\begin{equation}
\label{ta-2}
\frac{c}{4\pi}\frac{1}{2}(\mathbf{[DB]+[EH]})|_{\overline{\Gamma}; \mathbf{v}=0}=
\frac{\varepsilon\mu +1}{2}\frac{c}{4 \pi}[\mathbf{EH}]\,{.}
\end{equation}
In order to get here the Poynting vector (\ref{Poynting}),
in accordance with the requirement (\ref{e3-3}), one has to subtract from
(\ref{ta-2}) the following quantity
\begin{equation}
\label{ta-3}
\frac{\varepsilon\mu -1}{2}\frac{c}{4 \pi}[\mathbf{EH}]=
\frac{c}{4\pi}\frac{1}{2}(\mathbf{[DB]-[EH]})|_{\overline{\Gamma}; \mathbf{v}=0}
\,{.}
\end{equation}
In the co-moving frame, the spatial component of the 4-vector
\begin{equation}
\label{ta-4}
\frac{\varepsilon\mu -1}{2}\,
\frac{c}{4\pi}\,\Omega_\alpha\,{,}
\end{equation}
where $\Omega_\alpha$ is the Minkowski vector (\ref{Ruh}), (\ref{Ruh-1}),
reproduces the left-hand side of Eq.\ (\ref{ta-3}). Bearing this in mind,
it is natural to add to the tensor (\ref{ta-1}) the  symmetric tensor
constructed from two 4-vectors $u_\alpha$ and
$(\varepsilon\mu-1)\,\Omega_\beta$,
\begin{equation}
\label{ta-5}
T^{(1)}_{\alpha\beta}=\frac{1}{2}(\tm +T^{\text{M}}_{\beta \alpha})+\frac{1}{4\pi}\,
\frac{\varepsilon \mu-1}{2}(u_\alpha\Omega_\beta+u_\beta\Omega_\alpha)\,{.}
\end{equation}
Taking into account (\ref{e0}) and (\ref{Ruh-1}), one can easily verify
that the tensor (\ref{ta-5}) in the rest frame ($\mathbf{v}=0$) and under
the fulfillment the constitutive relations (\ref{e3-1})  satisfies the
conditions (\ref{e3-2})--(\ref{e3-4}).\footnote{The subtraction needed is
brought about due to the negative sign in front of
$\mathbf{\,S}^{\text{M}}$ and \(\mathbf{g}^{\text{M}}\) in~\eqref{e2-5}.}
The tensor~\eqref{ta-5} in view of~\eqref{Ruh-2} has vanishing trace.

Formally, the tensor \eqref{ta-5} is defined on the whole configuration
space $\Gamma$. However, in order to carry out the needed subtraction at
\(\mathbf{v}=0\) and fulfill condition \eqref{e3-3}, this tensor should be
considered only on $\g$. Obviously, this restriction must also be kept in
the case of moving medium ($\mathbf{v}\neq 0$), in order to have a smooth
dependence on \(\mathbf{v}\) of the tensor looking for.

Abraham proceeds exactly in this way. He defines the energy-momentum
tensor on $\g$ as the contraction of the tensor \( T^{(1)}_{\alpha\beta}
\) on $\g$~(see Ref.~\cite[p.~42]{AP1910}):
\begin{equation}
\label{ta-6}
 \ta\, \Bigl |_{\,\overline \Gamma}=\left . T^{(1)}_{\alpha\beta}\,
\right |_{\,\overline \Gamma}{.}
\end{equation}
The task to define the energy-momentum tensor on the whole
configuration space $\Gamma$ was not posed by Abraham.

By making use of Eqs.~\eqref{Minkowski-1}, \eqref{Minkowski-2}, we can
represent the 4-vector \((\varepsilon \mu -1)\,\Omega_\alpha\) on
$\overline \Gamma$ in the form~\cite[Eq.~(106)]{Kafka}
\begin{equation}
\label{ta-7}\left . (\varepsilon \mu
-1)\,\Omega_\alpha=u_\beta(D_\gamma B_{\alpha \beta
\gamma}-E_\gamma H_{\alpha \beta \gamma})\right |_{\overline \Gamma}\,{.}
\end{equation}
Further, it is worthwhile to use the property \eqref{Ruh-2} and  to
express the 4-vector \eqref{ta-7} in terms of one 3-dimensional vector
$\boldsymbol{\mathfrak{W}}$~(Eq.~(107) in Ref.~\cite{Kafka}),
\begin{equation}
\label{ta-8}
(\varepsilon \mu -1)\,\bm{\Omega}=\frac{\boldsymbol{\mathfrak{W}}}{\gamma}\,{,} \quad
(\varepsilon \mu -1)\,{\Omega}_4= \frac{\rmi}{\gamma}(\mathbf{q}\W)\,{.}
\end{equation}

In the case of isotropic medium moving with the velocity $\mathbf{v}$ all
the components of the energy-momentum tensor \eqref{ta-5},~\eqref{ta-6},
\begin{equation}
\label{a1} \ta=\left (
\begin{matrix}
\sigma^{\text{A}}_{ij}&-\rmi c\,
\mathbf{g}^{\text{A}}\cr -\frac{\displaystyle
\rmi}{\displaystyle  c}
\,\mathbf{S}^{\text{A}}&w^{\text{A}}\cr
\end{matrix}
\right ),
\end{equation}
were expressed by Abraham in terms of the field vectors
$\mathbf{E,\;H,\;D,\;B}$, the velocity of the medium  \(\mathbf{q}\), and
the vector $\W$~(Eqs.~($24_{a,b,c}$) in Ref.~\cite{AP1910})
\begin{gather}
\label{w1}\sigma^{\text{A}}_{ij}=\frac{1}{2}\,(\sigma^{\text{M}}_{ij}+\sigma^{\text{M}}_{ji})+
\frac{1}{8\pi}(q_i\mathfrak{W}_j+q_j\mathfrak{W}_i)\,{,} \\
\label{w2}
\frac{1}{c}\,\mathbf{S}^{\text{A}}=\frac{1}{8\pi}\,\{\mathbf{[EH]+[DB]}-\W-\mathbf{q}(\mathbf{q}
\W)\}
=c\mathbf{g}^{\text{A}}{,}\\
\label{w3}
w^{\text{A}}=w^{\text{M}}-\frac{1}{4\pi}\,(\mathbf{q}\W)\,{.}
\end{gather}
In turn, the vector \( \W\), defined in~\eqref{ta-8}, is also expressed in
terms of the field vectors and the velocity of the medium. Abraham
(Ref.~\cite{AP1910}, p.~42) uses here the constitutive relations in
the 3-dimensional vector form \eqref{Minkowski-3}, \eqref{Minkowski-4}.
Kafka applies here the 4-dimensional notation~(Ref.~\cite{Kafka},
pp.~48--51). As~a preliminary, Kafka proves the relation (Eq.~(119) in
Ref.~\cite{Kafka})
\begin{equation}
\label{w-4} (\varepsilon
\mu-1)(u_\alpha\Omega_{\beta}-u_\beta\Omega_{\alpha})\left
|_{\overline \Gamma}=F_{\alpha\gamma} H_{\gamma
\beta}-F_{\beta\gamma }H_{\gamma \alpha}\right |_{\overline
\Gamma}\,{,}
\end{equation}
which holds on $\g$.

We shall not reproduce here a rather cumbersome proof of this
formula \cite{Kafka} and only note the following. On the left-hand side of
Eq.~\eqref{w-4}, the dependence on \( \varepsilon \mu \) disappear as a
result of using  the constitutive relations \eqref{Minkowski-1},
\eqref{Minkowski-2}. When the pair of indices $\alpha\,\beta$ in
Eq.~\eqref{w-4} assumes the values 1\,4, 2\,4, 3\,4, the following
equality arises here:
\begin{equation}
\label{w-5}
-\rmi\{\W-\mathbf{q}(\mathbf{q}\W)\}=-\rmi\{\mathbf{[DB]-[EH]}\}\,{,}
\end{equation}
or in another form
\begin{equation}
\label{w-6}
\W=\mathbf{[DB]-[EH]}+\mathbf{q}(\mathbf{q}\W)\,{.}
\end{equation}
From  \eqref{w-6} it follows
\begin{eqnarray}
(\mathbf{q}\W)&=&(\mathbf{q,[DB]-[EH]})+q^2(\mathbf{q}\W)= \nonumber \\
&=& \gamma^2(\mathbf{q,[DB]-[EH]})\,{.}\label{w-7}
\end{eqnarray}
Finally, we arrive at the result
\begin{equation}
\label{w-8}
\W=\mathbf{[DB]-[EH]}+\gamma^2 \mathbf{q (q, [DB]-[EH])}\,{.}
\end{equation}

The substitution of \eqref{w-8} into  \eqref{w2} and  \eqref{w3}  yields
\begin{gather}
\frac{1}{c}\,\mathbf{S}^{\text{A}}=c\mathbf{g}^{\text{A}}=\frac{1}{4\pi}
[\mathbf{EH}]-\frac{\mathbf{q}}{4\pi(1-q^2)}\,(\mathbf{q},\mathbf{[DB]-[EH]}){,}\label{w-9}\\
w^{\text{A}}=\frac{1}{8\pi}
(
\mathbf{ED+HB})-\frac{1}{4\pi(1-q^2)}\,(\mathbf{q},\mathbf{[DB]-[EH]})\,{.}
\label{w-10}
\end{gather}

Here we shall not write out the bulky formulae which are obtained as a
result of substituting Eq.~\eqref{w-8} into~\eqref{w1}. Their explicit
form is obvious. The energy-momentum tensor constructed in this way is
symmetric.

Let us note once more that Eqs.~\eqref{w1}--\eqref{w3} and
\eqref{w-8}--\eqref{w-10} determine the components of the Abraham
energy-momentum only on $\g$ and do not include the material
characteristics of the medium. The Minkowski constitutive relations
\eqref{Minkowski-1}, \eqref{Minkowski-2} or \eqref{Minkowski-3},
\eqref{Minkowski-4} are used by Abraham specifically. Namely, they are not
used directly to express one pair of field vectors from the set
$\mathbf{E,\;H,\;D,\;B}$ in terms of the rest pair, but special formulae
are used which contain all the vectors $\mathbf{E,\;H,\;D,\;B,\;v}$ and
the material characteristics $\varepsilon,\;\mu$, these formulae being
valid only under  fulfillment of the Minkowski constitutive relations,
i.e., on~$\g$. An important example of such a formula is the equality
\eqref{w-4}. When the pair of indices $\alpha\,\beta$ in Eq.~\eqref{w-4}
takes the values 2\,3, 3\,1, and 1\,2, one obtains the
equality~\cite{Kafka}
\[
[\mathbf{q}\W]=\mathbf{[ED]+[HB]}\,{.}
\]
Taking into account \eqref{w-6}, we deduce
\begin{equation}
\label{w-10a}
\mathbf{[q,[DB]-[EH]]=[ED]+[HB]}\,{.}
\end{equation}
It is obvious that this equality holds only under
fulfillment of the Minkowski relations \eqref{Minkowski-3},
\eqref{Minkowski-4}.

Abraham deduced his formulae only on $\g$. Therefore, it is not clear
whether 10 quantities defined in this way can be definitely considered as
the components of a 4-dimensional tensor.\footnote{Appealing to the
works \cite{AP1910,Grammel,Kafka}, Abraham believed that the
energy-momentum tensor, constructed by him, possesses correct properties
under Lorentz transformations~\cite[p.~309]{Abraham1920}.}
In order to verify this one has, first of all, to define the quantities
under consideration on the whole configuration space~$\Gamma$. This  will
be done in Sec.~\ref{symt}.

Taking advantage of formula \eqref{w-4} we can represent the tensor \(
T^{(1)}_{\alpha\beta}\) \eqref{ta-5}, defined on $\g$, in another form.
Let us write the definition \eqref{ta-5} and formula \eqref{w-4},
multiplied by \(1/2 \), on the neighbouring lines,
\begin{align}
4\pi T^{(1)}_{\alpha
\beta} &=\frac12\,F_{\alpha \gamma}H_{\gamma \beta}+\frac12\,F_{\beta \gamma}H_{\gamma \alpha}
-\frac14\,\delta_{\alpha\beta}F_{\gamma \delta}H_{\delta\gamma}+\frac{\varepsilon\mu-1}{2}\,
(u_\alpha\Omega_\beta+u_\beta\Omega_\alpha)\,{,}\label{w-11}
\\
0&=\frac12\,F_{\alpha \gamma}H_{\gamma \beta}-\frac12\,F_{\beta \gamma}H_{\gamma \alpha}
\hbox to 32mm{\hfil$-$\hfil}
\frac{\varepsilon\mu-1}{2}\,
(u_\alpha\Omega_\beta-u_\beta\Omega_\alpha)\,{.}\label{w-12}
\end{align}
Adding and subtracting the left-hand sides and the right-hand sides of
these equations we obtain on~$\g$,
\begin{equation}
\label{w-13}
\ta=T^{(1)}_{\alpha\beta}=T^{(2)}_{\alpha\beta}=T^{(3)}_{\alpha\beta}\,{,}
\end{equation}
where
\begin{align}
T^{(2)}_{\alpha\beta}=\tm +\frac{\varepsilon\mu-1}{4\pi}\,u_\beta\Omega_\alpha\,{,}
\label{w-14}\\
T^{(3)}_{\alpha\beta}=T^{\text{M}}_{\beta\alpha}+\frac{\varepsilon\mu-1}
{4\pi}\,u_\alpha\Omega_\beta\,{.}
\label{w-15}
\end{align}
If we use in  \eqref{w-12} Eq.\ \eqref{w-4}, multiplied by an arbitrary
real number, then we  obviously can continue to infinity the  chain of
equalities in \eqref{w-13}, and consequently in the
definition~\eqref{ta-6}. However, the tensors \( T^{(i)}_{\alpha\beta}, \;
i=2,3,\dots\) constructed in this way give nothing new in comparison with
$T^{(1)}_{\alpha\beta}$. On $\g$, they are equal to $T^{(1)}_{\alpha\beta}$
and to the Abraham tensor according to \eqref{ta-6}, that is all these
tensors are simply another form of representation for the Abraham tensor.
In spite of the fact that these tensors represented in terms of
vector~$\W$ in accordance with~\eqref{ta-8}, do not, in general case,
coincide with Eqs.~\eqref{w1}--\eqref{w3} and are not explicitly
symmetric, but upon replacing~$\W$ by~\eqref{w-8}, the components of these
tensors are defined by \eqref{w-9}--\eqref{w-10} and Maxwell stress
tensors also coincide.

Let us show this using the tensor $T^{(2)}_{\alpha \beta}$, which is,
according to construction, the most close to the Minkowski energy-momentum
tensor. The components of the tensor $T^{(2)}_{\alpha \beta}$ are defined
by Eqs.~\eqref{ta-8} and~\eqref{w-14}:
\begin{gather}
\sigma^{(2)}_{ij}=\sigma^{\text{M}}_{ij}+\frac{1}{4\pi}\,\mathfrak{W}_i q_j\,{,}
\label{wt2-1}\\
\frac{1}{c}\,\mathbf{S}^{(2)}=\frac{1}{4\pi}\,\mathbf{[EH]}+
\frac{1}{4\pi}\,\mathbf{q}(\mathbf{q}\W)\,{,}
\label{wt2-2} \\
c\mathbf{g}^{(2)}=\frac{1}{4\pi}\,\mathbf{[DB]}-\frac{1}{4\pi}\,\W\,{,}
\label{wt2-3} \\
w^{(2)}=w^{\text{M}}-\frac{1}{4\pi}\,(\mathbf{q}\W)\,{.}
\label{wt2-4}
\end{gather}
It is this form that was used by Abraham for the representation  of the
components of his energy-momentum tensor in his book
\cite[\S39, Eqs.~(199e), (199d), (201), and
(201a)]{Abraham1920}. Replacing in~\eqref{wt2-2} and~\eqref{wt2-3} the vector~$\W$ by
\eqref{w-8}, we can easily obtain \eqref{w-9} and \eqref{w-10}. The
identity of the 3-dimensional Maxwell stress tensor represented in the
forms \eqref{wt2-1} and \eqref{w1} is proved by Abraham in his
article \cite[pp.~42,~43]{AP1910}.

Pauli in his survey  \cite[Eq.\ (303)]{Pauli}, does not note
that the tensors \( T^{(i)}_{\alpha\beta}, \; i=1,2,3\) are equal to each
other only on $\g$ and that in the rest frame these tensors coincide only
due to the constitutive relations \eqref{e3-1}. In view of this  the
status of Eq.\ (303) in the Pauli treatment \cite{Pauli} was left
unclear.\footnote{Pauli's assertion \cite{Pauli} ``The identity of three
expressions (303) follows from their coincidence in the co-moving
reference frame $K'\,$'' is simply wrong, being literally understood.}

The employment of the tensors $T^{(i)}_{\alpha \beta}, \; i=1,2,3$ outside
$\overline \Gamma$\cite{Leonhardt} is not physically
justified because outside $\g$ these tensors   have no  relation to the
Abraham energy-momentum tensor. Indeed, outside $\g$ these tensors are
different and already due to this reason one cannot consider them as an
extension of the Abraham formulae \eqref{w1}--\eqref{w3},
\eqref{w-8}--\eqref{w-10} on the whole configuration space~$\Gamma$.  The
consistent extension of the Abraham formulae on $\Gamma$ will be
implemented in  Sec.\ \ref{symt} of the present paper.

\section{The symmetric energy-momentum tensor}
\label{symt}

In the Introduction, it was explained that the term ``symmetric
energy-momentum tensor'' implies the following. This tensor is defined by
known values of its components in the rest frame
\cite[Eqs.~(298) and (300)]{Pauli}) and is determined on the whole configuration
space $\Gamma$. In the present Section we construct this tensor
generalizing the Abraham reasoning considered  in the preceding Sec.\ \ref{ta}.
This will enable us to prove, rather simple, the identity   of the
symmetric energy-momentum tensor and the Abraham tensor on~$\g$. Let us
recall  that the term ``Abraham tensor'' we kept for Eqs.\
\eqref{a1}--\eqref{w3}, \eqref{w-8}--\eqref{w-10} derived by Abraham
himself only on~$\g$.

It is obvious that conditions \eqref{e3-2}--\eqref{e3-4}
enable us  to define the symmetric energy-momentum tensor
on $\Gamma$ if we demand their fulfilment in the co-moving
reference frame without using the constitutive relations
\eqref{e3-1}. Below we demonstrate this.

In the generic inertial reference frame, tensor
\eqref{ta-1} yields the density of the energy flux,
\begin{equation}
\label{4e-1}
\frac{c}{4\pi}\frac{1}{2}(\mathbf{[DB]+[EH]})\,{.}
\end{equation}
In order to get here Poynting vector \eqref{Poynting},
in agreement with condition \eqref{e3-3}, one must subtract
from \eqref{4e-1} the following quantity:
\begin{equation}
\label{4e-2}
\frac{c}{4\pi}\frac{1}{2}(\mathbf{[DB]-[EH]})\,{.}
\end{equation}
Let us consider the 4-vector $\widetilde\Omega_\alpha$
which is obtained from the Minkowski vector $\Omega_\alpha$
\eqref{Ruh} by the substitution
\begin{equation}
\label{4e-3}
 \mathbf {E} \to \mathbf
{D},\quad \mathbf {H} \to \mathbf {B}, \quad
\mathbf {D} \to \mathbf {E}, \quad \mathbf {B}
\to \mathbf {H}\,{.}
\end{equation}
Using Eq.~\eqref{Ruh}, one can construct for $\wt_\alpha$ the
explicitly covariant formula,
\begin{equation}
\label{4e-4}
\wt_\alpha=B_{\alpha \beta \gamma}u_\beta D_\gamma\,{.}
\end{equation}
In the co-moving reference frame, Eq.~\eqref{4e-4} gives
\begin{equation}
\label{4e-5}
\frac{c}{4\pi}\,(\wt'_1,\wt'_2,\wt'_3)=\frac{c}{4\pi}\,\mathbf{[D'B']}\,{,}\quad
\wt'_4=0
\end{equation}
(compare with Eqs.\  \eqref{Ruh-1}, \eqref{Poynting}).

Now it is clear that the 4-vector \eqref{ta-7}, which was used by Abraham
in constructing the energy-momentum tensor only on $\g$, should be
replaced by the difference
\begin{gather}
\label{4e-6} \wt_\alpha-\Omega_\alpha = u_\beta(D_\gamma
B_{\alpha \beta \gamma}-E_\gamma H_{\alpha \beta
\gamma})=\\
=F_{\alpha \nu}H_{\nu \lambda}u_\lambda-H_{\alpha \nu}F_{\nu \lambda}u_\lambda=
\widetilde \omega_\alpha- \omega_\alpha\,{,}\label{4e-7}
\end{gather}
where
\begin{equation}
 \label{4e-8}
\omega_\alpha=H_{\alpha\nu}F_{\nu\lambda}u_\lambda,
\quad \widetilde \omega_\alpha=F_{\alpha\nu}H_{\nu\lambda}u_\lambda \,{.}
\end{equation}
In place of $T^{(1)}_{\alpha \beta}$ defined in
\eqref{ta-5}, we have now, on the whole configuration space
$\Gamma$, the tensor which we call, for definiteness, {\it the
symmetric energy-momentum tensor:}
\begin{equation}
\label{4e-9}
T^{\text{sym}}_{\alpha\beta}=\frac{1}{2}(\tm +T^{\text{M}}_{\beta \alpha})+
A_{\alpha \beta}\,{,}
\end{equation}
where
\begin{gather}
\label{4e-10}
A_{\alpha \beta}=A_{\beta \alpha}=\frac{1}{8\pi}\,
\{u_\alpha(\wt_\beta-\Omega_\beta)+u_\beta(\wt_\alpha-\Omega_\alpha)\}=\\
=\frac{1}{8\pi}\,\{
u_\alpha(F_{\beta \nu}H_{\nu \lambda}u_\lambda-H_{\beta\nu}F_{\nu\lambda}u_\lambda)+
u_\beta(F_{\alpha \nu}H_{\nu \lambda}u_\lambda-H_{\alpha\nu}F_{\nu\lambda}u_\lambda)
\}\,{.}\label{4e-11}
\end{gather}

Taking into account Eqs.\  \eqref{Ruh-1} and  \eqref{4e-5}, one can easily
verify that in the rest frame $(\mathbf{v}=0)$ the tensor
$T^{\text{\,sym}}_{\alpha\beta}$ satisfies the conditions
\eqref{e3-2}--\eqref{e3-4} without utilizing the constitutive relations
\eqref{e3-1}.

As was noted in the Introduction,  the
symmetric energy-momentum tensor \eqref{4e-9},
\eqref{4e-11}  was constructed in recent papers
\cite{Obukhov, MR} (the approach used in \cite{MR} is
discussed in Section \ref{f-ib} of the present paper). The
authors called it by the Abraham tensor but juxtaposition
of Eqs.\ \eqref{4e-9}, \eqref{4e-11} and Eqs.\
\eqref{w1}--\eqref{w3}, \eqref{w-8}--\eqref{w-10} was not
implemented in \cite{Obukhov,MR}. In our approach, the coincidence
of these tensors on $\g$ follows immediately from the
comparison of \eqref{4e-6} and~\eqref{ta-7}. In the
analytical way, it is proved as follows. Taking advantage
of relation \eqref{w-4}, valid on \(\g \) and being read
from right to left, and using property \eqref{Ruh-2} of the
Minkowski vector $\Omega_\alpha$, Equation \eqref{4e-11} can
be transformed to the form
\begin{equation}
\label{4e-11a}  A_{\alpha \beta}
\big |_{\g}=\frac{1}{4\pi}\,\frac{\varepsilon\mu-1}{2}\,(u_\alpha
\Omega_\beta+u_\beta\Omega_\alpha)\big |_{\g}\,{.}
\end{equation}
Keeping in mind Eqs.\  \eqref{4e-9}, \eqref{ta-5}, and
\eqref{ta-6}, we obtain
\begin{equation}
\label{4e-11b}
T^{\text{\,sym}}_{\alpha \beta}\big |_{\g}=T^{(1)}_{\alpha \beta}\big |_{\g}=
T^{\text{A}}_{\alpha \beta}\big |_{\g}\;{.}
\end{equation}
Below we prove equality  \eqref{4e-11b} once more by making use of the
3-dimensional vector formula~\eqref{w-10a} in place of~\eqref{w-4}.

When the energy-momentum tensor applies to practical calculations,  the
explicit form  of its components is, as a rule, needed in terms of the
3-dimensional vector notation. We turn now to the construction of such
formulae for the symmetric energy-momentum tensor on~$\Gamma$.

First we derive the 3-dimensional vector representation for the auxiliary
4-vectors~\(\omega_\alpha \) and~\(\widetilde\omega_\alpha \), introduced
in Eqs.~\eqref{4e-7} and~\eqref{4e-8}. From~(\ref{e0}) and~(\ref{Kafka-1})
it follows
\begin{equation}
\label{4e-11c}
\omega_\alpha=\gamma\{
[\mathbf{EH}]-\mathbf{[H[\mathbf{q}B]]}+\mathbf{D}(\mathbf{q}\mathbf{E});\;
\rmi(\mathbf{ED})-\rmi (\mathbf{q}[\mathbf{D}\mathbf{B}])\}\,{.}
\end{equation}
Obviously, the 4-vector  \( \widetilde \omega_\alpha\)
is derived from \( \omega_\alpha\) by substitution \eqref{4e-3},
\begin{equation}
\label{4e-12}
{\widetilde \omega}_\alpha=\gamma\{
[\mathbf{DB}]-\mathbf{[B[\mathbf{q}H]]}+\mathbf{E}(\mathbf{q}\mathbf{D});\;
\rmi(\mathbf{ED})-\rmi (\mathbf{q}[\mathbf{E}\mathbf{H}])\}\,{.}
\end{equation}
The difference ${\widetilde \omega}_\alpha- \omega_\alpha$ becomes
\begin{gather}
{\widetilde \omega}_\alpha- \omega_\alpha=\gamma \{
\mathbf{[DB]+E(qD)+H(qB)-[EH]-D(qE)-B(qH)};\;\nonumber \\
\mathbf{\rmi (q[DB])-\rmi(q[EH])
}\}\,{.} \label{4e-13}
\end{gather}

By analogy with \eqref{ta-8} it is convenient to represent
the difference \eqref{4e-6}, \eqref{4e-7}, and
\eqref{4e-13}  in the following form:
\begin{equation}
\label{4e-14}
\widetilde{\boldsymbol{\Omega}}-{\boldsymbol{\Omega}}=\frac{\mathbf{W}}{\gamma}\,{,}
\quad \wt_4-\Omega_4=\frac{\rmi}{\gamma}\,(\mathbf{qW})\,{,}
\end{equation}
where the vector $\mathbf{W}$ is
\begin{equation}
\label{4e-15}
\mathbf{W}=\gamma^2
\{\mathbf{[DB]-[EH]+E(qD)+H(qB)-B(qH)-D(qE)
}\}\,{.}
\end{equation}
Equation \eqref{4e-15} yields in particular
\begin{equation}
\label{4e-16}
\mathbf{(qW)}=\gamma^2
\{\mathbf{(q[DB])-(q[EH])}\}\,{.}
\end{equation}

Let us introduce the standard notation for the components
of the tensor $\tsym$ in a generic inertial reference frame
$K$, where the medium velocity is $\mathbf{v}$,
\begin{equation}
\label{tsym} \tsym=\left (
\begin{matrix}
\sigma^{\text{sym}}_{ij}&-\rmi c\,
\mathbf{g}^{\text{sym}}\cr -\frac{\displaystyle
\rmi}{\displaystyle c}
\,\mathbf{S}^{\text{sym}}&w^{\text{sym}}\cr
\end{matrix}
\right ){.}
\end{equation}
Equations~\eqref{4e-9}, \eqref{4e-10}, and~\eqref{4e-14} yield for the
individual components of the symmetric tensor~\eqref{tsym} the following
3-dimensional vector formulae:
\begin{gather}
\label{w1sym}\sigma^{\text{sym}}_{ij}=\frac{1}{2}\,(\sigma^{\text{M}}_{ij}+\sigma^{\text{M}}_{ji})+
\frac{1}{8\pi}(q_i W_j+q_j{W}_i)\,{,} \\
\label{w2sym}
\frac{1}{c}\,\mathbf{S}^{\text{sym}}=\frac{1}{8\pi}\,\{\mathbf{[EH]+[DB]}-\mathbf{W}
-\mathbf{q}(\mathbf{qW})\}
=c\mathbf{g}^{\text{sym}}{,}\\
\label{w3sym}
w^{\text{sym}}=w^{\text{M}}-\frac{1}{4\pi}\,(\mathbf{qW})\,{.}
\end{gather}
These formulae are also obtained by the substitution
$\W\to\mathbf{W}$ in analogous equations
\eqref{w1}--\eqref{w3}  for the Abraham tensor. Equations
\eqref{w1sym}--\eqref{w3sym} and \eqref{4e-15} determine,
in an explicit form, the symmetric energy-momentum tensor
\eqref{tsym} on the whole configuration space~$\Gamma$.

Comparing Eqs.~\eqref{w-7} and~\eqref{4e-16}, we infer that the energy
density given by the symmetric energy-momentum tensor \eqref{w3sym} and by
the Abraham tensor~\eqref{w3} are equal to each other on the whole
configuration space~$\Gamma$ and are defined by~\eqref{w-10},
\begin{equation}
\label{sym-a}
w^{\text{sym}}=w^{\text{A}}=\frac{1}{8\pi}
(
\mathbf{ED+HB})-\frac{1}{4\pi(1-q^2)}\,(\mathbf{q},\mathbf{[DB]-[EH]})\,{.}
\end{equation}
In the general case, the remaining components of these tensors, by virtue
of~\eqref{w-8} and~\eqref{4e-15}, do not coincide outside~$\g$. However,
when the problem under consideration allows to direct one of the
coordinate axes  (for example, the $x$-axis) along the medium velocity
\( \mathbf{v} \),
\begin{equation}
\label{sym-b}
\mathbf{q}=(q_x,0,0)\,{,}
\end{equation}
then some other components of these tensors will also equal
on $\Gamma$, namely, the diagonal components of the Maxwell
stress tensors,
\begin{equation}
\label{sym-c}
\sigma^{\text{sym}}_{xx}=\sigma^{\text{A}}_{xx},\quad
\sigma^{\text{sym}}_{yy}=\sigma^{\text{A}}_{yy},\quad
\sigma^{\text{sym}}_{zz}=\sigma^{\text{A}}_{zz}
\end{equation}
and the components
\begin{equation}
\label{sym-d}
\sigma^{\text{sym}}_{yz}=\sigma^{\text{sym}}_{zy}=
\sigma^{\text{A}}_{yz}=\sigma^{\text{A}}_{zy}\,{.}
\end{equation}
The first equality  in  \eqref{sym-c} is the consequence of
the relation
\begin{equation}
\label{sym-ã}
\mathbf{W}_x=\gamma^2\{\mathbf{[DB]}_x-\mathbf{[EH]}_x
\}=\W_x\,{,}
\end{equation}
which holds under condition \eqref{sym-b} (see Eqs.~\eqref{4e-15}
and~\eqref{w-8}). The remaining equalities in~\eqref{sym-c} and
equalities~\eqref{sym-d} follow from~\eqref{sym-b}.

Now we prove that the 3-dimensional vectors $\mathbf{W}$ \eqref{4e-15} and
$\W$ \eqref{w-8} equal on $\g$. For this purpose, we construct the vector
product of $\mathbf{q}$ and the relation \eqref{w-10a} valid on~$\g$,
\begin{gather}
\mathbf{[q[q\bigl ([DB]-[EH] \bigr )]]}=\mathbf{q(q,[DB]-[EH])}-q^2\mathbf{([DB]-[EH])}={}\nonumber\\
{}=\mathbf{[q,[ED]]+[q[HB]]}=\mathbf{E(qD)-D(qE)+H(qB)-B(qH)}\,{.} \label{4e-17}
\end{gather}
The substitution of \eqref{4e-17} into  \eqref{4e-15} yields
\begin{gather}
\mathbf{W}=\gamma^2\left\{\mathbf{[DB]-[EH]}-q^2\mathbf{([DB]-[EH]}+\mathbf{q}
\left (\mathbf{q},\mathbf{[DB]-[EH]}
\right )
\right \}={}\nonumber\\
{}=\mathbf{[DB]-[EH]}+\gamma^2\mathbf{q}(\mathbf{q,([DB]-[EH])})=\W\,{.}\label{4e-18}
\end{gather}
It implies that the symmetric energy-momentum tensor
\eqref{tsym}--\eqref{w3sym}  and the Abraham tensor
 \eqref{a1}--\eqref{w3} are identical on~$\g$. In other words, this
inference may be formulated like this: the correct extension of the
Abraham formulae \eqref{a1}--\eqref{w3} onto the whole configuration space
$\Gamma$ is accomplished by symmetric energy-momentum tensor
\eqref{tsym}--\eqref{w3sym}. Obviously, this extension is unique as the
tensor $\tsym$ is defined by its value in $K'$ \eqref{e3-2}--\eqref{e3-4}
uniquely.

Closing this section, we make an important remark. In practical
calculations, aimed at obtaining  the final physical result, it is
obligatory assumed to use  Minkowski constitutive relations
\eqref{Minkowski-1}, \eqref{Minkowski-2} or \eqref{Minkowski-3},
\eqref{Minkowski-4}. Therefore, in such studiesá it makes sense to utilize,
from the very beginning, the energy-momentum tensor in the Abraham form as
the respective formulae are substantially more compact in comparison with
those given by the symmetric tensor. This concerns the density of the
energy current, the linear momentum density~\eqref{w-9}, and especially
the components of the 3-dimensional Maxwell stress tensor in the
form~\eqref{wt2-1}.

\protect
\section{Form-invariance of the vector formulae for the energy-momentum tensor}
\label{f-i}
\protect
\subsection{The methods of defining the Minkowski tensor
and the form-invariance of the  vector formulae involved}
\label{f-ia}

In this ûubsection, we compare two methods of defining the 4-dimensional
tensor: in the explicitly covariant form and in the 3-dimensional vector
form. As an example, we shall use the Minkowski energy-momentum tensor. In
this case one can clearly show the covariance properties of the
3-dimensional vector formulae for the Minkowski tensor, which result in
their form-invariance.

\protect
\subsubsection{Definition of the Minkowski tensor by equation (\ref{e2-4})}
\label{f-iaa}

First we consider the definition of  the Minkowski tensor by
4-dimensional explicitly covariant expression (\ref{e2-4}), and, in
addition, we postulate that the latter is valid both for the medium at rest
and for the moving medium. It is worth emphasizing  that the velocity of
the medium does not enter Eq.\ (\ref{e2-4}). In terms of the 3-dimensional
vector notation the components of the tensor $\tm$ (\ref{e2-4}), in the
arbitrary inertial reference frame $K$, are given by Eqs.\
\eqref{e2-5}--\eqref{e8}.

It is obvious that all the components of the Minkowski
tensor (\ref{e2-5}), expressed in terms of the field
vectors $\mathbf{E,H,D,B}$ according to
Eqs.~(\ref{e2-6})--(\ref{e8}), keep their functional
dependence, specified by these formulae, in a generic
inertial reference frame. Indeed, when passing from
Eq.~(\ref{e2-4}) to Eqs.~(\ref{e2-5})--(\ref{e8}) the
reference frame was not fixed or specified. This assertion
is true both for the medium at rest and for the moving
medium since  the velocity of medium enter neither
explicitly covariant formula (\ref{e2-4})  nor the
consequent vector formulae (\ref{e2-5})--(\ref{e8}).

Thus the transformation of the tensor $\tm$, defined in a generic
reference frame $K$ originally by an explicitly covariant formula
(\ref{e2-4}) and afterwards written in the vector form
(\ref{e2-5})--(\ref{e8}),  to a new inertial reference frame, $K''$, is
simply reduced to the substitution $\mathbf{F \to F}''$ in Eqs.\
(\ref{e2-6})--(\ref{e8}),
\begin{equation}
\label{e9} T\,^{''\text{M}}_{\alpha
\beta}(\mathbf{F}'')=T\,^{\text{M}}_{\alpha
\beta}(\mathbf{F}'')\,{,}
\end{equation}
where $\mathbf{F}$ means the set of the field vectors,
\begin{equation}
\label{e9a}
\mathbf{F= \{E,\; H,\; D,\;B} \}\,{.}
\end{equation}
As before, the field vectors without prime belong  to an arbitrary
inertial reference frame~$K$ and the vectors with two primes pertain to
analogous frame~$K''$. This form-invariance property of the 3-dimensional
vector formulae (\ref{e2-5})--(\ref{e8}) is a straightforward consequence
of the relativistic covariance of definition (\ref{e2-4}).

Certainly, the numerical values of the individual components of the
tensor, calculated by Eqs.~(\ref{e2-5})--(\ref{e8}) at respective points
$r_\alpha=(x,y,z, \rmi ct)$ and $r''=(x'',y'', z'', \rmi ct'')$, vary in
accordance with the transformation law for the tensor
\begin{equation}
\label{e9b} T{\,}''^{\,\text{M}}_{\alpha
\beta}(\mathbf{F}''(r''_\gamma))=\Lambda_{\alpha \delta}
\Lambda_{\beta \rho}T^{\text{M}}_{\delta
\rho}(\mathbf{F}(r_\sigma))\,{,}
\end{equation}
where $\Lambda_{\alpha \beta}$ is the matrix of the Lorentz transformation
connecting the points $r''_\alpha$ and  $r_\beta$,
\begin{equation}
\label{e10} r''_\alpha=\Lambda_{\alpha \beta}
r_\beta\,{.}
\end{equation}
The  assertion above, concerning the Lorentz transformation of the
components of the tensor~$T^{\text{M}}_{\alpha \beta}$ in the vectorial
form, holds obviously for an arbitrary tensor of any rank.

\protect
\subsubsection{Definition of the Minkowski tensor by the vector equations
(\ref{e2-5})--(\ref{e8})} \label{f-ibb} The Minkowski energy-momentum
tensor, as a generic tensor, can be uniquely defined by fixing its
components in a specified reference frame. Let us assume that in the
co-moving frame $K'$, where the medium is at rest, the components of the
tensor $T{\,}'^{\,\text{M}}_{\alpha \beta}$ are defined by vector formulae
(\ref{e2-5})--(\ref{e8}). All the vectors and their components in these
formulae  should be provided with the prime, in accord with our notation.
In order to find the components of the tensor under consideration in an
arbitrary inertial reference frame we can proceed in two ways: (i) to
construct the 4-dimensional explicitly covariant formula, which reproduces
Eqs.\ (\ref{e2-5})--(\ref{e8}) in the reference frame $K'$, for example,
generalizing the energy-momentum tensor in the vacuum~\eqref{e2-1} and
\eqref{e2-4}; (ii) by making use of Eq.~\eqref{e9b} to transform the
3-dimensional Eqs.~(\ref{e2-5})--(\ref{e8}) to the generic reference
frame~$K$. Here we cannot take advantage of the form-invariance~(\ref{e9})
since Eqs.~(\ref{e2-5})--(\ref{e8}), according to the task posing, define
the Minkowski tensor not in an arbitrary reference frame~$K$, but in the
co-moving frame~$K'$.

The first way was, in fact, traced above (Section~\ref{f-iaa}). In~order
to apply here the Lorentz transformations one has to resort to
Eq.~(\ref{e9a}) which  in the case under consideration becomes
\begin{equation}
\label{e11} \tm(\mathbf{F}(r_\gamma))=\Lambda_{\alpha
\delta}\Lambda_{\beta
\varrho}T\,^{'\,\text{M}}_{\delta
\varrho}(\mathbf{F}(r'_\sigma))\,{.}
\end{equation}
According to the conditions of the task, the components of the
tensor~$T\,^{'\,\text{M}}_{\delta\varrho}(\mathbf{F}')$ in (\ref{e11}) are
given by the vectorial  equations (\ref{e2-5})--(\ref{e8}). Further one
has to express the field vectors~$\mathbf{F}'$ on the right-hand side of
Eq.~\eqref{e11} in terms of~$\mathbf{F}$. Having in mind that for
Eqs.~\eqref{e2-5}--\eqref{e8} there exist  4-dimensional explicitly
covariant representation \eqref{e2-4}, one can predict beforehand the
result of these transformations\footnote{The employment of the Lorentz
transformations in construction of the symmetric energy-momentum tensor is
discussed in Section~\ref{f-ib}.},
\begin{equation}
\label{e12}
\tm(\mathbf{F}(r_\gamma))=T\,^{'\,\text{M}}_{\alpha\beta}(\mathbf{F}(r_\gamma))\,{.}
\end{equation}
Nevertheless, in a recent paper \cite{Veselago-2}, these transformations
were carried out for the component~$T^{\text{M}}_{xx}$ and its
form-invariance was shown.\footnote{In Ref.\ \cite{Veselago-2} the
co-moving frame was denoted by $K$ and generic inertial frame by~$K'$. The
component $T^{\text{M}}_{xx}$, defined in~$K$ by Eq.~(\ref{e2-6}) (present
paper), was transformed into the reference frame  $K'$. It was shown that
$T^{'\text{M}}_{xx}(\mathbf{F}')=T^{\text{M}}_{xx}(\mathbf{F}')$.} This
property of the Minkowski tensor was treated in Ref.\ \cite{Veselago-2} not
completely correct, as the manifestation of {\it the relativistic
invariance} of this tensor (see critical note in \cite{MR}).
As~was explained above, this form-invariance  is in fact the consequence
of {\it the covariance property} of individual components of generic
tensor written {\it in vector notation}.

In pertinent literature \cite{LL2}, the form-invariance of the vectorial
formulae, determining the components of the 4-dimensional vectors and
tensors in the theory of electromagnetic field is not considered, probably
owing to clearness of this property. Nevertheless, we believe that
detailed discussion of this point  presented above will be useful.

\protect
\subsection{Form-invariance of the symmetric energy-momentum tensor
and the Abraham tensor in the vector notation}
\label{f-ib}

For these tensors, we have explicitly covariant
4-dimensional representations~\eqref{4e-9}, \eqref{4e-10}
and~\eqref{ta-5}, \eqref{ta-6}. Therefore,  3-dimensional
vectorial Eqs.~\eqref{tsym}--\eqref{w3sym}, \eqref{4e-15}
and~\eqref{a1}--\eqref{w3}, \eqref{w-8} or \eqref{w-9},
\eqref{w-10} derived from them preserve the form of
functional dependence on field vectors $\mathbf{F}$
\eqref{e9a} and velocity of medium $\mathbf{v}$ when
passing from one inertial reference frame to another. The
transformation rule  \eqref{e9} generalized to this case
reads
\begin{gather}
T^{''\text{\,sym}}_{\alpha \beta}(\mathbf{F'',v''})=T^{\text{\,sym}}_{\alpha \beta}(\mathbf{F'',v''})\,{,}
\label{e13} \\
T^{''\text{A}}_{\alpha \beta}(\mathbf{F'',v''})=T^{\text{A}}_{\alpha \beta}(\mathbf{F'',v''})\,.
\label{e14}
\end{gather}
This rule enables one to find the tensors $T^{''\text{\,sym}}_{\alpha
\beta}(\mathbf{F'',v''})$ and $T^{''\text{A}}_{\alpha
\beta}(\mathbf{F'',v''})$ in an generic inertial reference frame  $K''$,
provided the reference frame $K$ with tensors $T^{\text{\,sym}}_{\alpha
\beta}(\mathbf{F,v})$ and $ T^{\text{A}}_{\alpha \beta}(\mathbf{F,v})$ is
a reference frame of {\it a generic type}, in particular, $\mathbf{v}\neq
0$, i.e., $K$ is not the co-moving frame.\footnote{One cannot agree with
the assertion, \cite{Groot} that Abraham declined the form-invariance
principle when constructing his energy-momentum tensor.} Hence, the
transformation rules~\eqref{e13} and~\eqref{e14} do not allow one to find
the symmetric energy-momentum tensor and the Abraham tensor in arbitrary
reference frame $K''$ proceeding from Eqs.~\eqref{e3-2}--\eqref{e3-4} for
these tensors in the rest frame $K'$ with $\mathbf{v} = 0$ (without
allowance for the constitutive  relations \eqref{e3-1} and with this
allowance, respectively). Here the method must be used, which was employed
in Sections \ref{ta}  and \ref{symt}, or one has to resort to the Lorentz
transformations \eqref{e11}. The latter became in the case of the tensor $
\tsym$
\begin{equation}
\label{e15}
\tsym(\mathbf F(r_\gamma))=\Lambda_{\alpha \delta} \Lambda_{\beta\rho}T^{'\text{\,sym}}_{\delta \rho}
(\mathbf F'(r'_\sigma))\,{,}
\end{equation}
where $\Lambda_{ \alpha \beta} $ is the matrix of the Lorentz transformation
connecting the points $r_\alpha$ and  $r'_\delta$,
\begin{equation}
\label{e16}
r_\alpha=\Lambda_{\alpha \delta}r'_\delta\,{.}
\end{equation}
As before, the variables in an arbitrary inertial reference frame $K$ are
not marked  by prime, and primed variables are referred to the co-moving
frame $K'$. Obviously, the velocity of generic inertial reference frame
$K$ with respect to the rest frame $K'$ is $-\mathbf{v}$, where
$\mathbf{v}$ is the velocity of the medium in $K'$. Hence, the matrix of
the Lorentz transformation in~\eqref{e16} is defined by the velocity
$-\mathbf{v}$: $\Lambda_{\alpha \beta}=\Lambda_{\alpha
\beta}(\mathbf{-v})$.

The tensor  $ T^{'\text{\,sym}}_{\alpha \beta}(\mathbf{F}')$ in the rest
frame  $K'$ is specified by Eqs.\  \eqref{e3-2}--\eqref{e3-4}. Further the
field vectors $\mathbf{F}'$ on the right-hand side of Eq.~\eqref{e15} must
be expressed  in terms of the field vectors~$\mathbf{F}$ in the generic
inertial reference frame~$K$. The latter is implemented by the
transformation~$\Lambda(+\mathbf{v})$.

The Lorentz transformations, considered above, were carried out in
\cite{MR} and, what is more, for all the components of the
tensor $\tsym$ at once and for arbitrary direction of the medium velocity
with respect to the coordinate axes.

For a single component of the symmetric energy-momentum tensor
$T^{\text{\,sym}}_{xx}=\sigma^{\text{\,sym}}_{xx}$ the analogous
transformations were in fact performed in the work \cite{Veselago-2}, the
$x$-axis being directed along the velocity of the medium~$\mathbf{v}$,
\begin{equation}
\label{e17}
\mathbf{v}=(v_x,0,0)\,{.}
\end{equation}
The authors  of Ref.\  \cite{Veselago-2} erroneously
interpreted the obtained dependence of the component
$T^{\text{\,sym}}_{xx}$ on $\mathbf{v}$ as an indication
that this energy-momentum tensor ``is not relativistically
invariant''. As a matter of fact, in that paper the exact
expression was derived for the $xx$-component of the
symmetric energy-momentum tensor in a generic inertial
reference frame, where the medium has the velocity
$v_x/c=-\beta$. One can be easily convinced of this by
comparing Equation (23) in the work \cite{Veselago-2},
preliminarily changing the sign of $\beta$, with the
$xx$-component of the Abraham tensor in the form $
T^{(2)}_{\alpha \beta}$  \eqref{w-14},
\eqref{wt2-1}--\eqref{wt2-4}. This tensor, borrowed from
Ref.\ \cite{Abraham1920}, is presented in the Appendix 4 of
the survey \cite{Skob}  under the condition~\eqref{e17}.
Here one has to bear in mind the following. The diagonal
elements of the tensor~$T^{(2)}_{\alpha \beta}$,
Eq.~\eqref{w-14}, and of the tensor $\ta$,
Eqs.~\eqref{ta-6}, \eqref{w1}--\eqref{w3}, obviously
coincide on the whole configuration space $\Gamma$.
Further, when condition \eqref{sym-b} is valid, equalities
\eqref{sym-c} and \eqref{sym-d} hold. Thus,
$\sigma^{(2)}_{xx}= \sigma^{\text{\,sym}}_{xx}$ on
$\Gamma$. In Ref.~\cite[p.\ 1361, Russian edition]{MR} it
was explained in what way the setting of the problem in the
work \cite{Veselago-2} should be supplemented in order to
get the correct inference given above. However, this
inference was not formulated in Ref.~\cite{MR}.

\protect
\section{Conclusion}
\label{cncl}
Let us summarize briefly the results obtained in our paper.
\begin{enumerate}
\item The Abraham energy-momentum tensor
\eqref{a1}, \eqref{wt2-1}--\eqref{wt2-4} and the covariant
Grammel formulae \eqref{ta-6}, \eqref{ta-5},
\eqref{w-13}--\eqref{w-15} for this tensor are defined only
on $\g$, where they coincide and do not depend on the
material characteristics of the medium $ \varepsilon$ and~$\mu$.
\item Employment of the Abraham and Grammel formulae
outside $\g$ is not justified physically.
\item The generalization of the Abraham reasoning enables us
to construct the symmetric energy-momentum tensor on the whole configuration space $\Gamma$.
\item The symmetric energy-momentum tensor coincides with the Abraham tensor on $\g$
and provides the unique relativistically covariant extension
of the Abraham tensor on the whole configuration space  $\Gamma$.
\item In practical calculations, the energy-momentum tensor can be used in the Abraham form
which is more compact in comparison with the symmetric tensor.
\item The vectorial  3-dimensional formulae for the components
of the symmetric energy-momentum tensor have the same form
(functional dependence) in the arbitrary inertial reference
frame, i.e., these formulae are form-invariant. The same
statement is also true for the Abraham tensor.
\end{enumerate}

The physical basis of the Abraham tensor and the symmetric energy-momentum
tensor is the definition of the linear momentum density of electromagnetic
field in a medium, which is postulated, in the rest frame, in the form
\eqref{e3-3}. Just this definition is  the physical reason which
distinguishes the Abraham approach from the Minkowski tensor and other
versions of the electromagnetic energy-momentum tensor in the medium. The
symmetry property of the Abraham tensor \eqref{a1}--\eqref{w3} and of the
tensor \eqref{4e-9}--\eqref{4e-11}, as well as their explicit dependence
on the velocity of the medium, are, really, the consequences of the
condition~\eqref{e3-3}.

On the face of it, one may infer from~\eqref{4e-9} that these properties,
symmetry, and explicit dependence on the medium velocity, are implemented
independently of each other. Indeed, the symmetric tensor is already
produced by the first two terms in \eqref{4e-9},\footnote{The symmetrized
Minkowski energy-momentum tensor is used in practical
calculations \cite{Tucker-1}.} and to take into account the dependence on
the velocity of the medium, one has to add one more term
$A_{\alpha\beta}(u)=A_{\beta\alpha}(u)$. As a matter of fact, it is not
the case. Namely, the symmetrization alone, without introducing an
explicit dependence on $u_\alpha$, does  not enable one to meet on
$\Gamma$ conditions \eqref{e3-2}--\eqref{e3-4} determining the tensor
$\tsym$.

Not numerus references, concerning the Abraham tensor, are, practically
all, enumerated in the preset paper \cite{Pauli, Abraham1920, AP1909,
AP1910}. Unfortunately, in the textbooks the Abraham energy-momentum tensor
is, as a rule, not considered.

Until now, the problem of defining the energy-momentum tensor of
electromagnetic field in the medium has been considered substantially for
the medium at rest. However the complete solution of the problem in
question demands also determination of the dependence  of this tensor on the
medium velocity. One may hope that  precise and comprehensive
consideration  of the mathematical aspects relating to  the Abraham
approach will be helpful for these investigations.

\begin{acknowledgments}
VVN thanks I.~Brevik for useful advices on the subject under study,
A.A.~Rukhadze for providing the reference to the paper \cite{MR}, and
Yu.N.~Obukhov for references~\cite{Schmutzer,Obukhov}.
\end{acknowledgments}


\begin{thebibliography}{99}

\bibitem{Pauli}
W.~Pauli, {\it Theory of Relativity} (New York: Pergamon Press, 1958) \S\S~33,~35;
[W.~Pauli, {\it Relativit\"atstheorie} (Leipzig: Teubner, 1921) \S\S~33,~35].

\bibitem{Grammel} R.~Grammel, Ann.\ Physik {\bf 346}, 570 (1913).

\bibitem{Ginzburg-Ugarov} V.L.~Ginzburg and V.A.~Ugarov,
Sov.\ Phys.\ Usp.\ {\bf 19}, 94 (1976)
[Usp.\ Fiz.\ Nauk {\bf 118}, 175 (1976)].

\bibitem{Leonhardt} U.~Leonhardt, Phys.\ Rev.\ A {\bf 73}, 032108 (2006).
\bibitem{Skob} D.V.~Skobel'tsyn, Sov.\ Phys.\ Usp.\ {\bf 16}, 381 (1973)
[Usp.\ Fiz.\ Nauk {\bf 110}, 253 (1973)].

\bibitem{Brevik-1} I.~Brevik, Mat.\ Phys.\ Medd.\ Dan.\ Vid.\ Selsk.\ {\bf 37},
No.~11, 1 (1970); {\bf 37}, No.~13, 1 (1970).

\bibitem{Brevik-2} I.~Brevik, Phys.\ Rep.\ {\bf 52}, 133 (1979).

\bibitem{Ginzburg-73} V.L.~Ginzburg, Sov.\ Phys.\ Usp.\ {\bf 16}, 434 (1973)
[Usp.\ Fiz.\ Nauk {\bf 110}, 309 (1973)].

\bibitem{Abraham1920}
M.~Abraham, {\it Theorie der Elektrizit\"at}, Bd.\ 2, 4.\ Aufl.\
(Leipzig: Teubner, 1920).

\bibitem{AP1909} M.~Abraham, Palermo Rend.\ {\bf 28}, 1 (1909);
[http://en.wikisource.org/wiki/Author:Max\underline{~}Abraham]

\bibitem{AP1910} M.~Abraham, Palermo Rend.\ {\bf 30}, 33 (1910);
[http://en.wikisource.org/wiki/Author:Max\underline{~}Abraham]

\bibitem{Schmutzer} E.~Schmutzer, {\it Relativistische Physik}
(Leipzig: Akademie Verlag, 1968) p.\ 408.

\bibitem{Groot} S.R.~de~Groot and L.G.~Suttorp,
{\it Foundations of Electrodynamics}
(Amstredam: North-Holland, 1972), Ch.~V, \S7.

\bibitem{Obukhov} Yu.N.~Obukhov, Ann.\ Phys.\ (Berlin), {\bf 17}, No.~9/10,
830 (2008).

\bibitem{MR} V.P.~Makarov and A.A.~Rukhadze, Phys.\ Usp.\ {\bf 52}, 937 (2011)
[Usp.\ Fiz.\ Nauk {\bf 181}, 1357 (2011)].

\bibitem{ni} J.\ Gratus, Yu.N.\ Obukhov, and R.N. Tucker,
% Conservation laws and stress-energy-momentum
% tensors for systems with background fields,
Annals Phys.
{\bf 327}, 2560 (2012).
% 2560 arXiv.org:1206.6704

\bibitem{nii}
    T.\ Ramos, G.F.\  Guillermo, Yu.N.\ Obukhov,
Phys.\ Lett.\ A {\bf  375},  1703 (2011).
% Ramos et al., Relativistic analysis of the dielectric
% Einstein box: Abraham, Minkowski and total energy-momentum tensors,
% pp. 1703-1709 arXiv.org:1103.1654

\bibitem{niii}
    T.\ Ramos, G.F.\  Guillermo, Yu.N.\ Obukhov,
% Ramos et al., First principles approach to the
% Abraham-Minkowski controversy for the momentum of light in
% general linear non-dispersive media,
J.\ Opt.\ {\bf 17}, 025611 (2015).
% arXiv.org:1310.0518

\bibitem{LL8} L.D.~Landau and E.M.~Lifshitz,
{\it Electrodynamics of Continuous Media} (Oxford: Pergamon Press, 1984), \S76.

\bibitem{LL2} L.D.~Landau and E.M.~Lifshitz,
{\it The Classical Theory of Fields} (Oxford: Pergamon Press, 1980).

\bibitem{Kafka} H.~Kafka, Ann.\ Physik {\bf 58}, 1 (1919).

\bibitem{Veselago-2} V.G.~Veselago and V.V.~Shchavlev, Phys.\ Usp.\
{\bf 53}, 317 (2010) [Usp.\ Fiz.\ Nauk {\bf 180}, 331 (2010)].

\bibitem{Tucker-1} S.~Goto, R.W.~Tucker, and T.J.~Walton,
Proc.\ R.\ Soc.\ A {\bf 467}, 59 (2011); {\bf 467}, 79 (2011).

\end{thebibliography}
\end{document}